\documentclass[12pt]{iopart}
\usepackage{graphicx}

\usepackage{graphicx}

\begin{document}

\title[Turbulent Kinetic Energy Spectra of Solar Convection]{Turbulent Kinetic Energy Spectra of Solar Convection from NST Observations and Realistic MHD Simulations}

\author{I.N. Kitiashvili$^{1-3}$, V.I. Abramenko$^{4}$, P.R. Goode$^{4}$, A.G. Kosovichev$^1$, S.~K. Lele$^{2,5}$, N.N. Mansour$^5$, A.A. Wray$^5$ and V.~B. Yurchyshyn$^{4}$}
\address{$^1$W.W. Hansen Experimental Physics Laboratory, Stanford University, Stanford, CA 94305, USA}
\address{$^2$Center for Turbulence Research, Stanford University, Stanford, CA 94305, USA}
\address{$^3$Kazan Federal University, Kazan, 420008, Russia}
\address{$^4$Big Bear Solar Observatory of New Jersey Institute of Technology,CA 40386, USA}
\address{$^5$Aeronautics and Astronautics Dept., Stanford University, Stanford, CA 94305, USA}
\address{$^6$NASA Ames Research Center, Moffett Field, Mountain View, CA 94040, USA}
\ead{irinasun@stanford.edu}

\begin{abstract}
Turbulent properties of the quiet Sun represent the basic state of surface conditions, and a background for various processes of solar activity. Therefore understanding of properties and dynamics of this `basic' state is important for investigation of more complex phenomena, formation and development of observed phenomena in the photosphere and atmosphere. For characterization of the turbulent properties we compare kinetic energy spectra on granular and sub-granular scales obtained from infrared TiO observations with the New Solar Telescope (Big Bear Solar Observatory) and from 3D radiative MHD numerical simulations ('SolarBox' code). We find that the numerical simulations require a high spatial resolution with 10 - 25~km grid-step in order to reproduce the inertial (Kolmogorov) turbulence range. The observational data require an averaging procedure to remove noise and potential instrumental artifacts. The resulting kinetic energy spectra show a good agreement between the simulations and observations, opening new perspectives for detailed joint analysis of more complex turbulent phenomena on the Sun, and possibly on other stars. In addition, using the simulations and observations we investigate effects of background magnetic field, which is concentrated in self-organized complicated structures in intergranular lanes, and find an increase of the small-scale turbulence energy and its decrease at larger scales due to magnetic field effects.
\end{abstract}
\maketitle

\section{Introduction}

Understanding and characterization of turbulent solar convection is a key problem of heliophysics and astrophysics. The solar turbulence driven by convective energy transport determines the dynamical state of the solar plasma, leads to excitation of acoustic waves \cite{kiti2011}, formation of magnetic structures \cite{Brand2011,kiti2010b} and other dynamical phenomena. Realistic numerical simulations of solar magnetoconvection are an important tool for understanding many observed phenomena, verification and validation of theoretical models, and interpretations of observations. The simulations of this type were started in pioneering works by Stein and Nordlund \cite{stein1998}, with the main idea of constructing numerical models based on first physical principles. The `quiet Sun' describes a background state of the solar surface layers without sunspots and active regions, that is, without large-scale magnetic flux emergence and other strong magnetic field effects, which can significantly change properties of the turbulent convection. Quiet-Sun regions are characterized by weak mean magnetic field of 1 - 10~G, which is usually concentrated in small-scale flux tubes in the intergranular lanes, and observed as bright points in molecular absorption lines. Previous investigations of the solar turbulent spectra from observations were presented by Abramenko {\it al.}
\cite{abramenko2001}, Goode {\it al.} \cite{goode2010a}, Matsumoto and Kitai \cite{Matsumoto2010}, Rieutord {\it al.} \cite{Rieutord2010}, Stenflo \cite{Stenflo2012} and others. Comparison of observations with numerical simulation data initially done by Stein and Nordlund \cite{stein1998} showed a good agreement between correlation power spectra obtained from smoothed simulation data and high-resolution observations from the La Palma. Such comparison of results of realistic-type MHD modeling with high-resolution observations gives us an effective way to understanding observed phenomena. Recently, advanced computational capabilities made it possible to construct numerical models of the solar turbulent convection with a high level of realism. On the other hand, modern high-resolution observational instruments with adaptive optics, such as the 1.6-m New Solar Telescope at the Big Bear Solar Observatory \cite{goode2010b} have allowed us to capture small-scale dynamics of the surface turbulence \cite{abramenko2011,goode2010a,yurch2011}.

 In this paper, we compare the turbulent kinetic energy spectra from observed and simulated data sets for the conditions of quiet-Sun regions, and investigate properties of solar turbulence and background magnetic field effects. We use two types of data: 1) high-resolution observations of horizontal flows from the New Solar Telescope (NST/BBSO, \cite{goode2010a}), and 2) high-resolution 3D radiative MHD and hydrodynamic simulations \cite{kiti2012}.

 \begin{figure}
 \centerline{\includegraphics[width=1\textwidth]{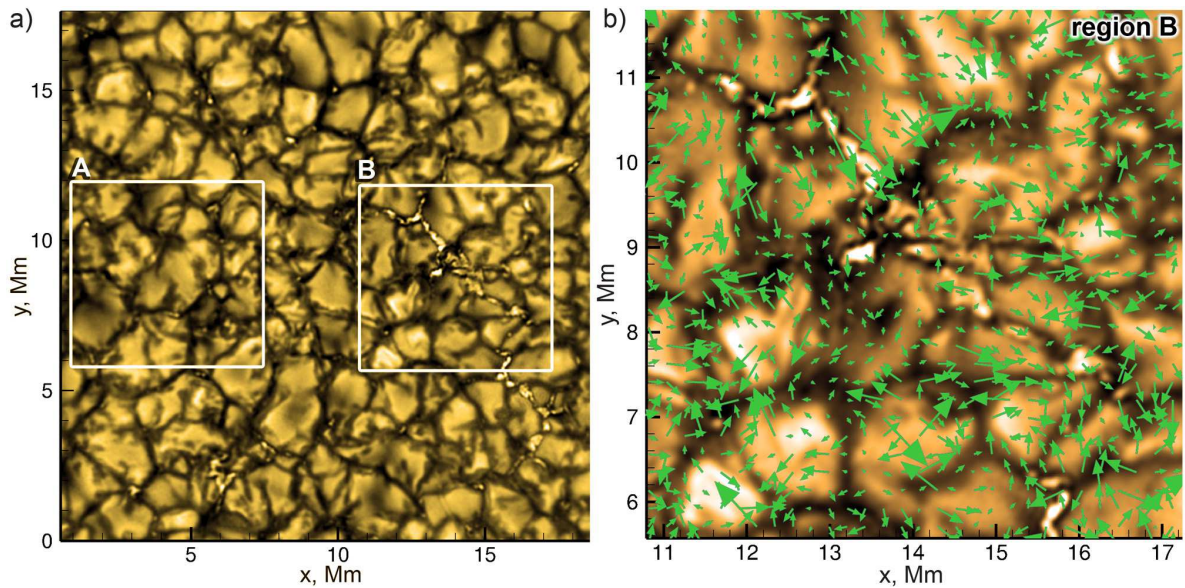}}
 \caption{A quiet-Sun region observed in the TiO filter with the New Solar Telescope (NST) on August 3, 2010. Squares in panel {\it a}) show two subregions: subregion $A$ without magnetic bright points, and subregion $B$ with conglomerates of magnetic bright points concentrated in the intergranular lanes. In panel {\it b}), subregion $B$ is shown in detail with overploted velocity field derived by a Local Correlation Tracking (LCT) method.}\label{fig:TiO}
 \end{figure}

\section{Observational data}
For the comparison we use broadband TiO filter (centered at 7057{\AA}) data of a quiet-Sun region obtained with the New Solar Telescope at Big Bear Solar Observatory  (NST/BBSO, \cite{goode2010b}) on August 3, 2010. The telescope has a 1.6-m aperture (with an off-axis design) and an adaptive optics system, implementing a speckle image reconstruction \cite{woger2008}, which allows to achive the diffraction limited resolution of $\sim 77$~km in this spectral range. The image sampling is $0.0375''$ ($\sim 27$~km) per pixel. The unprecedented spatial resolution together with the high temporal resolution, 10~s allows us to resolve and investigate the structure and dynamics of very tiny structures on the Sun, such as jet-like structures on the scale of a granule or less \cite{goode2010a}, substructure of granules \cite{yurch2011}, dynamics of magnetic bright points \cite{abramenko2010,Manso2011}, and turbulent diffusion properties of the solar convection \cite{abramenko2011}.

 The analyzed data-set of the quiet-Sun region with the size of $28.2''\times 26.2''$ includes a 2-hour time-sequence of TiO images with 10~sec cadence. To investigate how magnetic field affects the turbulent properties we select two subregions, marked as $A$ and $B$ in Figure~\ref{fig:TiO}{\it a}. Region $A$ has almost no magnetic bright points (BPs), whereas region $B$ includes conglomerates of BPs, which represent concentrations of magnetic field. A correlation between BPs and magnetic field structures was previously discussed by Berger and Title \cite{Berger2001}. For reconstruction of the horizontal velocity field from the observations, a Local Correlation Tracking method \cite{November1988,Strous1996} was used. Figure~\ref{fig:TiO}{\it b} shows an example of the velocity field plotted over the corresponding TiO intensity image for region $B$. Calculations of the energy spectra for both observational and simulation data sets were done by using the same code adopted in \cite{abramenko2001}.

\section{Numerical simulations}
\subsection{Radiative MHD code and computational setup}
For modeling, we use a 3D radiative MHD code (`SolarBox') developed for realistic simulations of top layers of the convective zone and lower atmosphere \cite{jacoutot08a}. The code takes into account the realistic equation of state, ionization and excitation of all abundant spices. Radiative energy transfer between fluid elements is calculated with a 3D multi-spectral-bin method, assuming the local thermodynamic equilibrium and  using the OPAL opacity tables \cite{Rogers1996}.  Initialization of the simulation runs is done from parameters of a standard model of the solar interior \cite{chris1996}. The sub-grid scale turbulence is modeled using a Large-Eddy Simulation (LES) approach \cite{Germano1991,Balarac2010}.
The simulations in this paper were obtained using a Smagorinsky eddy-viscosity model \cite{Smagorinsky1963}, in which the compressible Reynolds stresses are described by equations given by Moin {\it et al.} \cite{Moin1991} and Jacoutot {\it et al.} \cite{jacoutot08a}, with the Smagorinsky coefficients  $C_S=C_C=0.001$.

For investigation of magnetic field effects, we impose a 10~G initially uniform vertical magnetic field. This field gets concentrated in compact flux-tube like structures in intergranular lanes and mimics magnetic field in the solar bright points. In all cases, the simulation results were obtained for a computational domain of $6.4\times6.4\times6.2$~Mm$^3$, including a 1~Mm high layer of the atmosphere, with a grid spacing of $\Delta x=\Delta y=12.5$~km and $\Delta z=10$~km. The lateral boundary conditions are periodic. The top boundary is open to mass, momentum, and energy transfers and also to the radiative flux. The bottom boundary is open for radiation and flows, and simulates energy input from the interior of the Sun.

\subsection{Effects of the spatial resolution}
One critically important issue of investigation of turbulent properties of convection is limited spatial resolution. In observations this means not resolving small-scale information. In numerical simulations, unresolved small-scale dynamics can affect general turbulent properties of convection due to missing physics of turbulent dissipation. The LES models of turbulence effectively increase the Reynolds number and capture, in part, dynamics on sub-grid scales, thus providing a more realistic representation of turbulent convection. An important requirements for the LES models is resolving all essential scales of convection, including the transition to the inertial (Kolmogorov) range \cite{Frisch1995}.

 \begin{figure}
 \centerline{\includegraphics[width=0.9\textwidth]{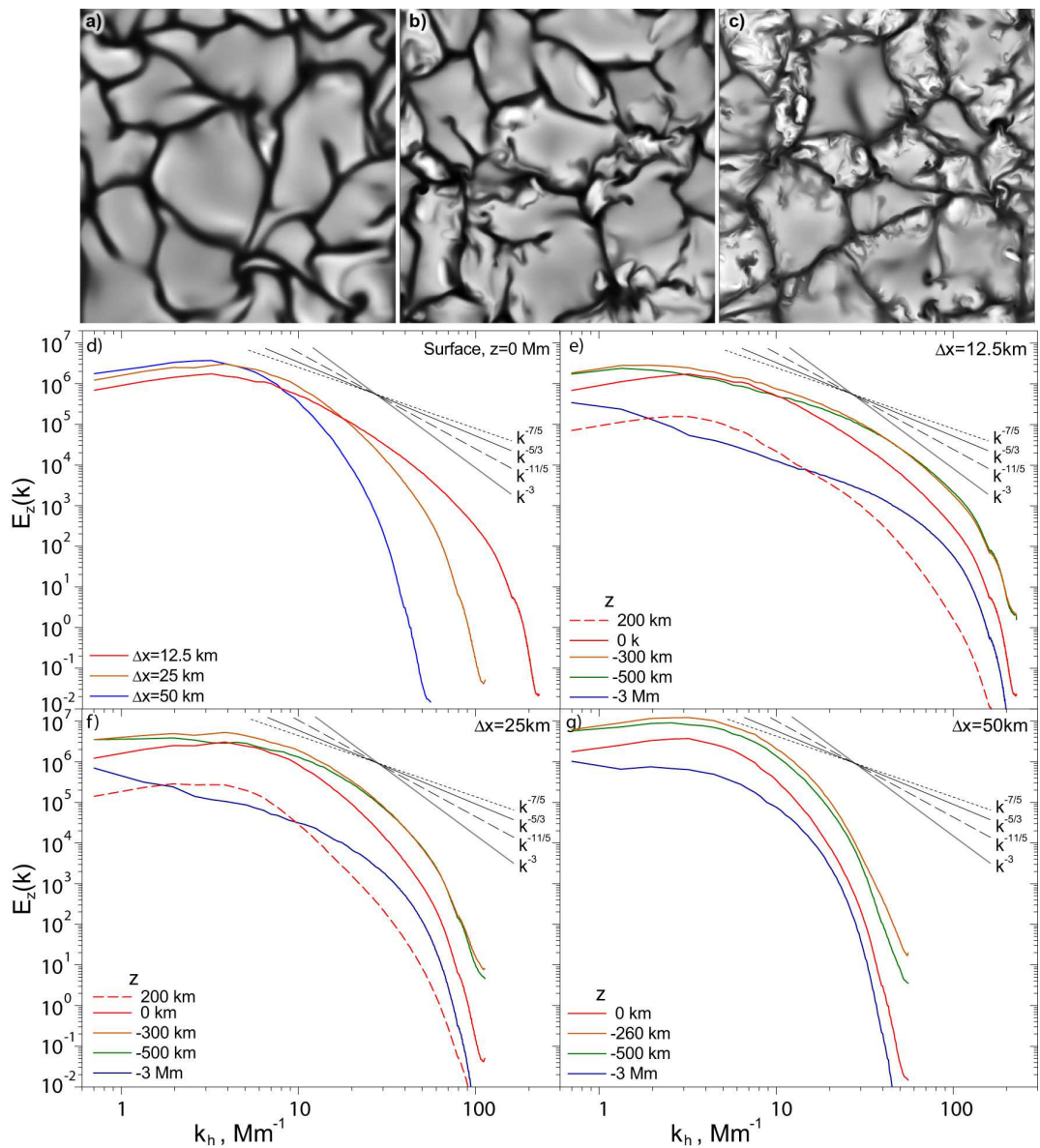}}
 \caption{Effect of numerical resolution on properties of the simulated convection. Top panels show surface snapshots of the vertical velocity for 50~km (panel {\it a}), 25~km ({\it b}) and 12.5~km (panel {\it c}) horizontal resolution. Panels {\it d-g} illustrate effects of the different numerical resolution on the turbulent energy spectra of the vertical velocity: at the photosphere layer (panel {\it d}), and at different locations above and below the photosphere (panels {\it e-g}).} \label{fig:B0sp}
 \end{figure}

Figure~\ref{fig:B0sp} shows the effect of numerical resolution on properties of turbulent vertical velocity spectra in our hydrodynamic simulations of solar convection for three cases of the horizontal grid-spacing: 50, 25 and 12.5~km. It is not surprising that the simulations with higher resolution reveal numerous, inhomogeneously distributed small-scale flow substructures, mostly concentrated at granular edges, and also more complicated dynamics of granules (panels {\it a-c}). The resolution effect is critical from the point of view of the energy cascade, because unresolved substructures may cause redistribution of energy through all scales.
For example, comparison of the turbulent spectra in the photosphere layer for the different resolutions (Fig.~\ref{fig:B0sp}{\it d}) shows a faster energy decay for large wavenumbers (small scales) and slightly higher energy density values on larger scales for the low resolution (50~km) case (blue curve). Such dependence of the power density slope on the resolution and the effect of the energy increase at large scales previously was found by  Stein and Nordlund \cite{stein1998}. In the high-resolution simulation spectrum (12.5~km, red curve), the inertial and dissipative subranges, expected from turbulence theories (e.g. \cite{Frisch1995}), can be identified. Because of the strong density stratification the spectral properties change with depth below the surface, and also change above the surface. The power density spectra for the horizontal resolutions of 12.5 and 25~km (Fig.~\ref{fig:B0sp}{\it e-f}) show similar variations of the turbulent properties of convection at different depths. The layers above the solar surface are characterized by a smaller total energy and a higher spectral energy density slope. These layers are convectively stable, and the turbulence spectrum reflects convective overshooting. The subsurface layers have stronger, more energetic motions, but the energy density slope decreases. In the deeper layers due to the decreasing velocity magnitude the kinetic energy decreases. Also, in the deeper layers the turbulent scales become larger, flows are more homogeneous; and the energy spectra can be described by the Kolmogorov ($-5/3$) power law \cite{Kolmogorov1941}. The low resolution simulations (50~km, Fig.~\ref{fig:B0sp}{\it g}) are capable to capture only the magnitude of the kinetic energy, but unlike the high-resolution simulations do not show the differences of the turbulent dynamics in different layers.

\section{Power spectra and data averaging}
For the numerically simulated convection (which in this case is modeled from first principles including all most significant physics contributions) it is important to resolve all essential scales, including the inertial subrange. Once the inertial subrange is resoled it is assumed that the turbulent cascade will continue to the dissipative subrange following the Kolmogorov law scaling. Following the Reynolds's idea of separation of turbulent flows on mean and fluctuating parts, we consider smoothly evolving averaged flows \cite{Monin1963}.
Because properties of the averaging can affect the resulting power spectra \cite{goode2010a}, we consider the energy spectra without averaging and with 3 different types of averaging (Table~\ref{table}), where case 1 represents an ensemble time-averaging with an overlapping averaging window, and in cases 2 and 3, we divide the whole data set in to individual temporal bins, 2 and 5 min long.

\begin{table}[b]
\caption{Parameters of time-averaging.}\label{table}
  \begin{tabular}{|c|c|c|l|}
    \hline
      & $T_w$, s & $T_s$, s & Comments \\
    \hline
      0 & -- & -- & no averaging \\
      1 & 20 & 10 & windows overlapping \\
      2 & 120 & 120 & average by bins \\
      3 & 300 & 300 & average by bins \\
    \hline
\end{tabular}
\end{table}

\begin{figure}
\centerline{\includegraphics[width=1\textwidth]{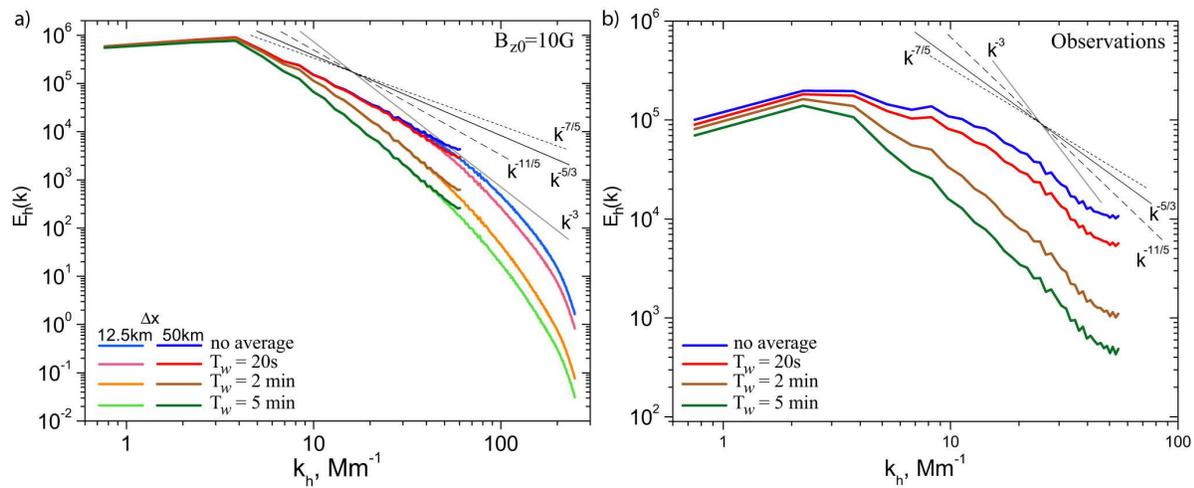}}
\caption{Effect of temporal averaging on the energy spectral density for the horizontal velocity fields in the simulations with an initial vertical field strength 10~G (panel {\it a}), and the horizontal velocities reconstructed by LCT from the TiO observations at NST/BBSO (panel {\it b}). Panel {\it a} shows also deviations of the energy spectra for the full resolution of the numerical data (12.5~km) and for the resolution degraded to the observational resolution (50~km).  Each panel shows the kinetic energy spectra for the original data set without averaging (blue curves), and for the filtered data sets obtained using the sliding averaging (black and green curves), and 2-min and 5-min bin averaging (yellow and red-brown curves).} \label{fig:aver}
\end{figure}

Figure~\ref{fig:aver} shows the influence of the different-type averaging on the kinetic energy spectra for the simulations with the initial magnetic field strength of 10~G (panel {\it a}), and for the observational data (panel {\it b}). For both, the numerical model and observational data, the averaging shows similar effects, in particular, an increase of the energy spectra slope. This corresponds to  stronger energy filtering of the smaller scales. For the  simulated data degraded to the observed spatial resolution, the difference in the energy spectra from the original high resolution (12.5~km) simulations appears only at the smallest resolved scales, due to the turbulent energy cascade cut off at the smaller unresolved scales. In the observational data such increasing of energy also takes place. In analysis of the solar turbulent dynamics, we would like to keep the maximum amount of the observed signal, therefore we use ensemble averaging with minimal filtering window properties (case 1, Table~\ref{table}; \cite{Reynolds1895}).

\begin{figure}
\centerline{\includegraphics[width=1\textwidth]{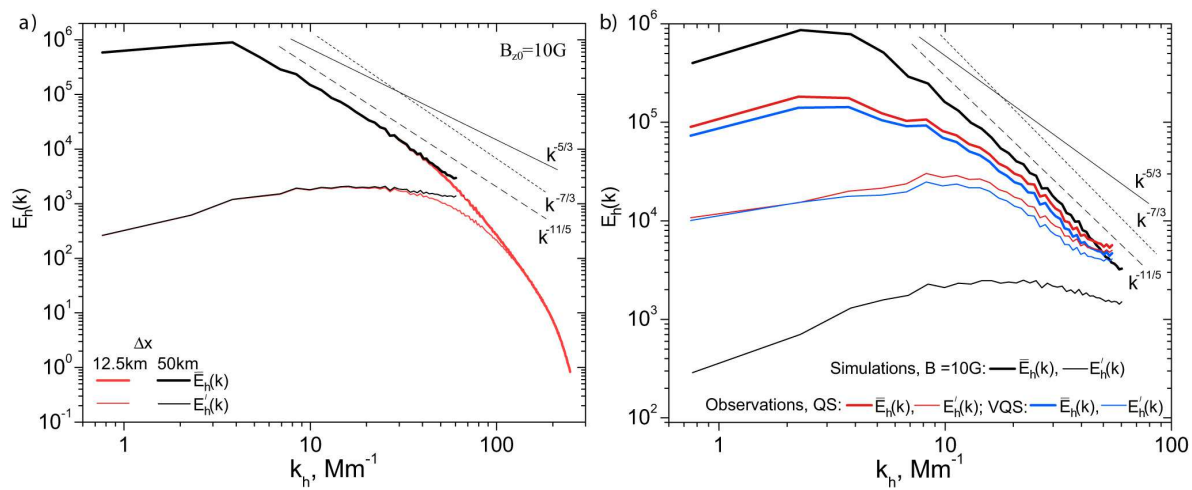}}
\caption{Comparison of the mean and fluctuating parts of the horizontal kinetic energy density obtained by the ensemble averaging from simulated and observed data sets (case 1). Panel {\it a}) shows comparison of the power spectra of the horizontal energy for the mean (thick curves) and fluctuating (thin curves) parts in the high-resolution simulations, $\Delta x=12.5$~km (red curves), and the simulation data with the degraded spatial resolution ($\Delta x=50$~km, black curves) for the weak mean initial magnetic field, $B_{z0}=10$~G. Panel {\it b}) shows the energy density spectra for the degraded simulated data (black curves) and for the quiet-Sun subregion $B$ with magnetic bright points (QS), and for subregion $A$ without magnetic bright points (VQS), indicated in Fig.~\ref{fig:TiO}{\it a}, region $A$).} \label{fig:averS}
\end{figure}

Because we would like to keep most of the signal we use a minimal possible averaging window ($T_w=20$~s, for 10~s cadence data series) with a corresponding window  time shift, $T_s=10$~s. Thus, in this case, the averaging of two closest in time frames causes filtration of fluctuations with time-scale less than 20~sec.
Fig.~\ref{fig:averS} illustrates the spectra for the mean (thick curve) and fluctuating (thin) parts of the horizontal velocities for the simulated and observed data obtained by the ensemble averaging. Panel {\it a} shows the energy density spectra obtained from the MHD simulations (with 10~G mean field, red curves). To see the effects of the spatial resolution we degraded the resolution of the simulated data to the resolution of  observed data ($\sim 50$~km, black curves). Because there is no noise in the simulated data, the spectra obtained from the degraded data follow the full resolution spectra on larger scales. The deviations become noticeable only at the smallest resolved scales.

\section{Discussion}
\begin{figure}
\centerline{\includegraphics[width=0.95\textwidth]{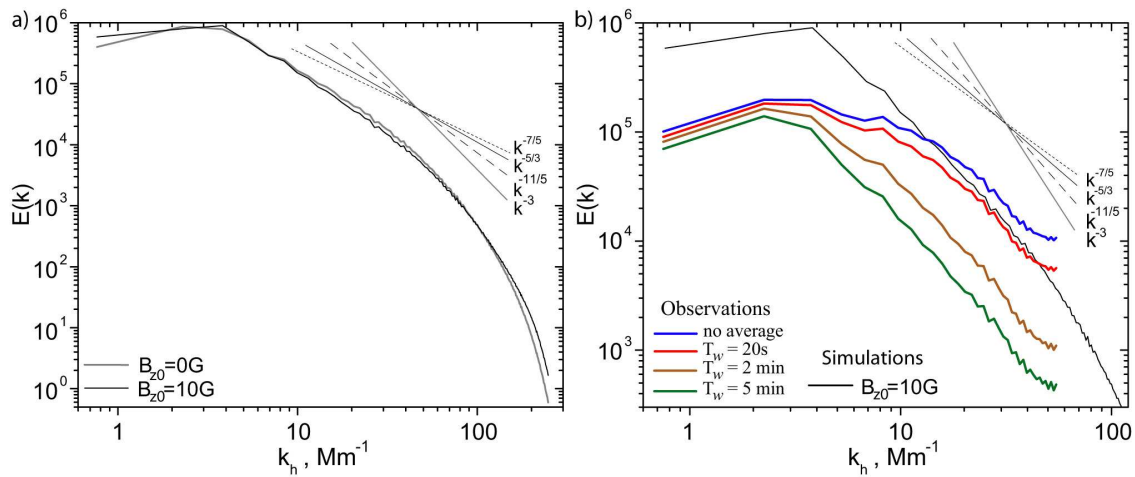}}
\caption{Effect of the background magnetic field (panel {\it a}) and comparison of the kinetic energy spectra for the simulations (with $B_{z0}=10$~G) and observations using the horizontal flow velocities reconstructed by the Local Correlation Tracking method from NST/BBSO observations (panel {\it b}).} \label{fig:comp}
\end{figure}

Investigation of solar convection is interesting from the point of view of the hydrodynamic turbulent properties of the highly stratified medium, and also for understanding and characterization effects of background magnetic fields on the turbulent energy transport between different scales. Recent numerical simulations have shown that the presence of a weak magnetic field can increase the level of nonlinearity and have different effect at large and small scales due to the increasing inhomogeneity of convective properties and magnetic coupling of plasma motions. In particular, decreasing of the turbulent kinetic energy at the sub-granular scales in the simulations with magnetic field (black curve, Fig.~\ref{fig:comp}{\it a}) can be caused by local suppression of turbulent motions near convective granular edges, where the magnetic field is collapsed in to small-scale concentrations of magnetic field ($\sim 1$~kG). Recent investigation of quiet-Sun data from the Hinode space mission showed strong, relatively high contribution of the collapsed field in the magnetic energy density distribution with a maximum at 80~km scale, and increasing of the magnetic energy density on granule scales (see histogram at Fig.~8 in \cite{Stenflo2012}).
Thus, the comparison of the energy density spectra for the hydrodynamic and weakly magnetized convection at the solar surface in Figure~\ref{fig:comp}{\it a} shows a higher kinetic density energy on scales less than 50~km in the presence of magnetic field, and opposite on larger scales. Actually, a similar effect of the collapsed magnetic flux was found by Stenflo \cite{Stenflo2012}, but with an exponential decrease of the energy density on small scales. Thus, on the small scales (less than 50~km) the increase of the kinetic energy density reflects an interplay of the collapsing flux dynamics and, probably, a small-scale dynamo action.  Perhaps, the  increase of the kinetic energy density on the small scales contributes to quasi-periodic flow ejections into the solar atmosphere by the small-scale vortex tubes as discussed by Kitiashvili {\it et al.} \cite{kiti2012}. This potential relationship needs to be investigated.

As discussed early, the data averaging allows us to filter out noise, and make data sets more homogeneous. However, increasing of the averaging window size can also filter out short-living features, and cause smearing of granules. Therefore, the averaging effect on the energy spectra, when most of energy on smallest scales is filtered, mostly leads to a steeper energy spectra slope. Averaging over two and more minutes makes the slope of the energy spectrum corresponding to the Kolmogorov power law ($k^{-5/3}$, \cite{Kolmogorov1941}). Such behavior of the  energy spectra reflects the famous in the turbulence literature Landau's  `Kazan remark' \cite{Frisch1995}, in which Landau draw attention to the absence of localized small-scale turbulent fluctuations in the Kolmogorov theory. Only when such fluctuations are filtered the spectrum becomes of the Kolmogorov type, as this happens in our case.

Comparison of the kinetic energy spectra calculated from the observational and simulated data sets shows a higher contribution of  flows with small wavenumbers in the simulations than in the observed data (Fig.~\ref{fig:comp}{\it b}). The extra power in the simulated data on these scales can come from the geometry of our numerical setup, in which convection is confined in a box with periodic boundary conditions in lateral directions, which can cause cutting of the energy transfer to larger-scale convective modes (e.g. due to inverse cascades). Also, this deviation can be caused by an underestimation of the velocity magnitude due to a degrading spatial resolution of the LCT (Local Correlation Tracking) data analysis procedure.

We have also analyzed effects of the averaging procedure with different parameters (Table~1) on the resulting power spectra. The ensemble averaging with a minimal size window ($T_w=20$~s) and window shift ($T_s=10$~s, for 10~s cadence data) filters out most of the noise signal, and shows good qualitative and quantitative agreements with the simulated high-resolution data on the scales less than $\sim 150$~km (Fig.~\ref{fig:comp}{\it b}). In terms of general properties of the energy spectra, the time-averaging in short bins (2 or 5 min) shows good qualitative agreement with the spectral profile obtained from the simulated data for all scales resolved in the observations. Such good qualitative agreement of the kinetic energy  spectra between the simulations and filtered observational data can be also due to removal of additional observational artifacts (such as local uncorrelated deformations of images and other instrumental effects), which can have time scales up to several minutes. The comparison of the energy spectra observed on the small scales (with wavenumbers larger than 30~Mm$^{-1}$) with the spectra calculated from the simulation data degraded to the observational resolution shows at all cases an increase of the energy density.

For investigation of magnetic field effects, we compared the kinetic energy spectra for the two selected regions (Fig.~\ref{fig:TiO}{\it a}), one of which (region $B$) was filled by magnetic bright points, and another (region $A$) almost did not have these features. Comparison of the energy spectra of these regions shows their almost identical behavior with the total energy smaller for region $B$. Because the difference between the both spectra is mainly in the energy magnitude, we can conclude that there was no significant difference in the turbulent dynamics.

Because the background magnetic field is present on the Sun everywhere, in order to get a more clear identification of magnetic effects we compare the spectra from the hydrodynamic and weakly magnetized surface turbulence simulations, and can see changes of the energy balance on different scales due to magnetic effects, namely: suppression of turbulent motions on the granular scales caused by the accumulation of magnetic field concentrations in the intergranular lanes, and, increasing of the kinetic energy density for large wavenumbers, probably, due to the small-scale dynamo action (Fig.~\ref{fig:comp}{\it a}).

\section{Summary}
We presented a comparison of the kinetic energy spectra of the solar turbulent convection obtained from the observed (NST/BBSO) and simulated ('SolarBox' code) horizontal velocity fields. Our analysis of the energy density spectra for different conditions of convective flows (with and without background magnetic field), different spatial resolutions and data averaging procedures found the following properties:
\begin{enumerate}
  \item The numerical simulations show good qualitative agreement with the observations in terms of the turbulence properties when the observational data a averaged in two-minute bins. This filtering removes from the observational data noise and relatively long-living ($\sim 1-2$~min) artifacts on spatial scales larger than the granule size. In order to reproduce the inertial (Kolmogorov) subrange it is necessary that the numerical simulations have sufficiently high spatial resolution, 10 - 25~km per grid step. In this case the transition from the inertial to the dissipative subrange is resolved, the Large Eddy Simulation (LES) turbulence modeling is justified.
  \item The ensemble averaging method is capable to filter most of noise in the data, and provided good qualitative and quantitative agreement between the observed and simulated turbulent spectra on the scales 300~km and less (Figs.~\ref{fig:averS}{\it b} and \ref{fig:comp}{\it b}).
  \item Different properties of the ensemble averaging (Table~1) used for the noise filtering cause changes in the energy spectra, leading in particular, to increasing of the slope of the spectra, both in the simulations and observations (due to stronger filtering on small scales, Fig.~\ref{fig:aver}).
  \item Degrading the simulation data to the spatial resolution of observations causes an increase of the  energy density on the smallest resolved scales (Figs.~\ref{fig:aver},~\ref{fig:averS});
\item The weak background magnetic field changes the energy balance on the different scales, namely: a) suppression of convective motions on larger scales due the magnetic field structures collapsed  in the intergranular lanes and restricting granule motions, and b) increasing of the kinetic energy density on the small-scales less than 50~km, probably due to a local small-scale dynamo action (Fig.~\ref{fig:comp}{\it a});
  \item The energy spectra change qualitatively with depth/height: in the deeper layers convective turbulence becomes more homogeneous and shows good correspondence to the Kolmogorov power-law turbulent energy cascade (Fig.~\ref{fig:B0sp});
\end{enumerate}

The good agreement between the observed and simulated spectra of the quiet-Sun convection opens perspectives for future detailed comparison between numerical models and observations. Our results show the importance of synergy between high-resolution observations and modern realistic-type MHD numerical simulations for understanding complicated turbulent phenomena on the Sun, in the direction of joint data analysis, interpretation and links between observations and models.

{\bf Acknowledgements.}
This work was partially supported by the NASA grant NNX10AC55G, the International Space Science Institute (Bern) and Nordita (Stockholm).

\end{document}